\documentclass[useAMS,usenatbib]{mn2e}
\usepackage{ifpdf} 
\ifpdf
  \pdfoutput=1  
  \usepackage[pdftex]{graphicx}         

  \usepackage[colorlinks, bookmarks, breaklinks, pdftitle={Recovering the intrinsic shape of early-type galaxies},pdfauthor={Remco C. E. van den Bosch and Glenn van de Ven}]{hyperref}
  \hypersetup{linkcolor=black,citecolor=black,filecolor=black,urlcolor=blue}       
  \usepackage[kerning]{microtype}
\else                                
  \usepackage[dvips]{graphicx} 
\fi

\usepackage[T1]{fontenc}
\usepackage{times}           
\usepackage[fleqn]{amsmath}
\usepackage{amssymb}  
\usepackage{journalnames}

\defcitealias{2008MNRAS.385..647V}{vdB08}
\defcitealias{2008MNRAS.385..614V}{vdV08}

\newcommand{\colfigref}[1]{\ref{#1}}
             
\newcommand{\kms}{km\,s$^{-1}$}
\newcommand{\dgr}{$^\circ$}
\newcommand{\Msun}{\textsc{m}$_\odot$}

\newcommand{\MLsun}{\textsc{m}$_\odot$/\textsc{l}$_\odot$}
 
\newcommand{\arcs}{\hbox{$^{\prime\prime}$}}

\newcommand{\ML}{\textsc{m}/\textsc{l}}

\newcommand{\eps}{\varepsilon}  
\newcommand{\NGCN}[1]{\textsc{ngc}\,#1}

\newcommand{\Sauron}{\textsc{sauron}}

\title[Intrinsic shape of early-type galaxies] {Recovering the
  intrinsic shape of early-type galaxies}

\author[van den Bosch \& van de Ven]{%
  Remco C.~E. van den Bosch$^{1,2}$, %
  Glenn van de Ven$^{3}$\thanks{Hubble Fellow}\\
  $^1$ McDonald Observatory, The
  University of Texas at Austin, 
  Austin, TX 78712, USA \ifpdf
  \href{mailto:bosch@astro.as.utexas.edu}{    
  [bosch@astro.as.utexas.edu]} 
  \else
  [bosch@astro.as.utexas.edu]
  \fi \\
  $^2$ Sterrewacht Leiden, Universiteit Leiden, Postbus 9513, 
       2300 RA Leiden,   The Netherlands \\
  $^3$ Institute for Advanced Study, Einstein Drive, 
       Princeton, NJ 08540, USA \ifpdf
    \href{mailto:glenn@ias.edu}{    
    [glenn@ias.edu]} 
    \else
    [glenn@ias.edu]
    \fi }
   
\date{Accepted 2009 June 1.  Received 2009 May 27; in original form 2008 November 20}

\pagerange{\pageref{firstpage}--\pageref{lastpage}} \pubyear{2009}
\begin{document}
\maketitle
\label{firstpage}

\begin{abstract}

  We investigate how well the intrinsic shape of early-type galaxies can be
  recovered when both photometric and two-dimensional stellar kinematic
  observations are available. We simulate these observations with galaxy
  models that are representative of observed oblate fast-rotator to triaxial
  slow-rotator early-type galaxies. By fitting realistic triaxial dynamical
  models to these simulated observations, we recover the intrinsic shape (and
  mass-to-light ratio), without making additional (ad-hoc) assumptions on the
  orientation.

  For (near) axisymmetric galaxies the dynamical modelling can strongly
  exclude triaxiality, but the regular kinematics do not further tighten the
  constraint on the intrinsic flattening significantly, so that the
  inclination is nearly unconstrained above the photometric lower limit even
  with two-dimensional stellar kinematics. Triaxial galaxies can have
  additional complexity in both the observed photometry and kinematics, such
  as twists and (central) kinematically decoupled components, which allows the
  intrinsic shape to be accurately recovered. For galaxies that are very round
  or show no significant rotation, recovery of the shape is degenerate, unless
  additional constraints such as from a thin disk are available.
  
   \end{abstract}

\begin{keywords} galaxies: elliptical and lenticular, cD - galaxies:
kinematics and dynamics - galaxies: structure \end{keywords}

\section{Introduction}
\label{sec:SRintro}

Numerical cold dark matter simulations predict that the dark matter haloes of
galaxies typically have intrinsic shapes with three distinct axes $a,b$ and
$c$ \citep[e.g.][]{2002ApJ...574..538J}. Such a triaxial shape is
characterized by the axis ratios $p=b/a$ and $q=c/a$ and $0 \le q \le p \le
1$, which can vary as a function of radius. Based on the specific angular
momentum inferred from integral-field stellar kinematics, early-type galaxies
seem to be divided into two classes of fast and slow rotators, different from
their morphological classification in ellipticals and lenticulars
\citep{1996ApJ...464L.119K,2007MNRAS.379..401E}. Most fast rotators, including
lenticular as well as many elliptical galaxies, which show strong rotation
around their photometric minor axis, seem consistent with oblate axisymmetry
($p=1$). However, many of these systems contain bars and also weakly triaxial
shapes can not be ruled out. On the other hand, the slow rotators, with little
or no rotation, are most likely triaxial.

However, all this is based on observed quantities, i.e., projected
onto the plane of the sky, while for the true comparison between
galaxies intrinsic properties, and thus the viewing directions are
needed. Most early-type galaxies lack thin discs (in either dust, gas
or stars), so that typically there is only a weak constraint on the
viewing direction from (mainly) the observed ellipticity.

Determining the intrinsic shape of galaxies is possible by statistical
analysis of their observed ellipticities \citep*[e.g.][]{1992MNRAS.258..404L,
1992ApJ...396..445R}. The deficiency of E0 type galaxies immediately rules out
that elliptical are perfectly axisymmetric
\citep*{1991ApJ...383..112F,1996ApJ...461..146R}, even if bars are taken into
account. A recent study by \cite{2008MNRAS.388.1321P} using ellipticities from
SDSS show that early-type galaxies are weakly triaxial with mean axial ratios
of $p\sim0.95$ and $q\sim0.7$. However, these kind of analyses involve rather
strong assumptions on the intrinsic shape distribution, and can not determine
the shape of individual galaxies. Alternatively, by assuming a functional form
for the intrinsic density it is possible to constrain the intrinsic shape of
the light distribution of an individual galaxy, given its observed ellipticity
and isophotal twist profile
\citep[e.g.][]{1981ApJ...244..458W,2008MNRAS.383.1477C}. In general, however,
there is a well-known non-uniqueness in the deprojection of the surface
brightness ---even for axisymmetric galaxies not viewed perpendicular to their
symmetry axis--- also called the `cone of ignorance'
\citep{1987IAUS..127..397R}. This tells us that it is not possible to
constrain the intrinsic shape using only the photometry without making strong
assumptions.

Nowadays, in addition to observations of the surface brightness of early-type
galaxies, integral-field spectrographs provide the full line-of-sight velocity
distribution (LOSVD) as a function of position on the sky
$\mathcal{L}(x',y',v_{z'})$. The LOSVD is commonly parameterized in terms of
its velocity moments, resulting in maps of the mean line-of-sight velocity,
velocity dispersion and higher-order velocity moments. Can we (better)
constrain the intrinsic shape, and hence the viewing direction, of galaxies
from these two-dimensional photometric \emph{and} kinematic observations?

Because the deprojection is degenerate and the conversion from light to mass
is complicated when dark matter is present, implying a range of possible
gravitational potentials. Even ignoring (or parameterizing) both these
aspects, it is far from evident that the LOSVD provides enough information to
infer the intrinsic shape. Let us first consider three-integral oblate
axisymmetric models, i.e., with a distribution function (DF) that is a
function $f(E,L_z,I_3)$ of the energy $E$, the angular momentum component
$L_z$ parallel to the symmetry $z$-axis, and a non-classical third integral of
motion $I_3$. In the special case that a galaxy happens to be well
approximated by a two-integral DF $f(E,L_z)$, the density $\rho(R,z)$ (in the
cylindrical coordinates $R$ and $z$) uniquely determines the even part of
$f(E,L_z)$ and the mean streaming $\rho\langle v_\phi \rangle$ in the
meridional plane fixes the part of $f(E,L_z)$ that is odd in $L_z$
\citep{1986PhR...133..217D}. If both $\rho(R,z)$ and $\rho\langle v_\phi
\rangle$ are known, they define a two-integral DF completely. The observed
velocity dispersion and higher order velocity moments of the LOSVD then
provide additional information, which for example can be used to constrain the
inclination \citep[see e.g.][]{2006MNRAS.366.1126C}. This two-integral
semi-isotropic result was recently generalized by \cite{2008MNRAS.390...71C}
using a special class of anisotropic models in which the velocity ellipsoid is
assumed to be aligned with cylindrical coordinates, and the anisotropy
$\beta_z$ \citep[as defined by][]{1982MNRAS.200..361B} to be positive and
constant within a galaxy. Both assumption are based on an simplified
interpretation of fig.~1 in \cite{2006MNRAS.366.1126C}, and is at most a
first-order approximation.

In the realistic case of a three-integral DF $f(E,L_z,I_3)$, it might well be
that the full three-variable LOSVD $\mathcal{L}(x',y',v_{z'})$ is needed to
determine the DF, so that there is no information left to constrain the
inclination. Using long-slit spectroscopy it was generally impossible to
constrain the inclination \citep[e.g.][]{1998ApJ...493..613V,
2000AJ....119.1157G}. Initially, the availability of two-dimensional stellar
kinematics seem to allow a constraint on the inclination beyond the lower
limit from the photometry, even in the case of full three-integral
axisymmetric orbit-based Schwarzschild models \citep{2002MNRAS.335..517V}.
However, \cite{2005MNRAS.357.1113K} convincingly showed that, after carefully
taking into account numerical effects, similar three-integral axisymmetric
Schwarzschild models at different inclinations above the photometric limit can
(nearly) all fit the observed LOSVD within the measurement uncertainties.

\cite{2007MNRAS.381.1672T} showed that axisymmetric modeling can have
signficant systematics in the recovered mass-to-light ratios by modelling
triaxial merger remnants with their axisymmetric machinery. This shows that
triaxial modelling is strongly preferred over axisymmetric modelling, as real
elliptical galaxies are at least weakly triaxial.

In the triaxial case, the DF is again a function of three integrals of motion
$f(E,I_2,I_3)$, but the orbital structure in these models is substantially
richer than in the oblate axisymmetric models, with four major orbit families,
instead of only one. This introduces a ``fundamental'' non-uniqueness in the
recovery of the DF: whereas in the oblate axisymmetric case $\rho(R,z)$
uniquely defines the even part of $f(E,L_z)$, in the (separable) triaxial case
the density $\rho(x,y,z)$ does \emph{not} uniquely determine the even part of
$f(E,I_2,I_3)$, even though both of these are functions of three variables
\citep{1992ApJ...389...79H}. It is (yet) unknown how much specification of
$\mathcal{L}(x',y',v_{z'})$ can narrow down the range of possible DFs further.
Therefore, in the triaxial case it seems to be even harder, if not impossible,
to constrain the viewing direction.

However, while mathematically there are an infinite number of deprojections of
the surface brightness possible already for oblate axisymmetric systems viewed
away from edge-on, in practice only a limited range of deprojections lead to
intrinsic density distributions that are considered realistic for galaxies.
Similarly, we have shown in \cite*[hereafter
\citetalias{2008MNRAS.385..614V}]{2008MNRAS.385..614V} that the DF in the
triaxial case is in practice well recovered at the correct viewing direction
using orbit based models. At the same time, we found in \cite[hereafter
\citetalias{2008MNRAS.385..647V}]{2008MNRAS.385..647V} that if there is enough
complexity in the photometry and/or kinematics it is likely that even the
intrinsic triaxial shape is well constrained. Such complexities are for
example photometric and kinematic twists, misalignment between the photometric
minor axis and kinematic rotation axis (kinematic misalignment) and so-called
kinematically decoupled components (KDCs). For example, the velocity field of
NGC\,4365 \citep{2001ApJ...548L..33D} shows rotation around the photometric
short-axis in the core as well as around the long-axis in the outer parts.
While the central KDC excludes a prolate shape, the global long-axis rotation
is not possible in an oblate system, leaving a triaxial
intrinsic shape as the only solution (see also vdB08).

Statistical analysis of observed ellipticities and kinematic
misalignments can provide an estimate of the intrinsic shape
distribution \citep{1985MNRAS.212..767B, 1991ApJ...383..112F}.
However, this statistical approach requires a large sample of galaxies
for which the kinematic misalignment is accurately measured, which is
not (yet) available. \cite{1993A&A...275...61T} explored the use of
the observed ellipticity and kinematic misalignment to constrain the
intrinsic shape of individual galaxies, adopting a specfic form for
the intrinsic density and streaming motion. \citet[and references
therein]{1994AJ....108..111S} developed a velocity-field fitting
method for individual galaxies based on a solution of the continuity
equation, assuming a similar flow which separates the density and
streaming motion in a radial and angular part. Because this approach
is fast the large parameter space can be sampled finely enough to find
the most probable solution using Bayesian statistics. However, the
method assumes a (plausible) solution for the continuity equation, is
not self-consistent, and fits the first velocity moment only. By
comparison, the orbit-based Schwarzschild models that we employ allow
for general density distributions, are self-consistent, and fit the
full LOSVD. As mentioned above, even with such a general method making
optimal use of the state-of-the-art two-dimensional photometric and
kinematic observations, it is not evident how well the intrinsic shape
can be recovered.

In this paper, we use a set of realistic (nearly) axisymmetric and
triaxial galaxy models with a three-integral DF, to show that an
increasing complexity in the corresponding two-dimensional observables
indeed leads to a better recovery of the intrinsic shape. In
\S~\ref{sec:SRabelmodels}, we construct these Abel models based on the
description of \citetalias{2008MNRAS.385..614V}. We fit the
corresponding observables using the triaxial Schwarzschild method of
\citetalias{2008MNRAS.385..647V}, of which we highlight the most
relevant aspects in \S~\ref{sec:SRthe_modeling}.  We investigate the
resulting constraints on the intrinsic shape in
\S~\ref{sec:results_of_the_dynamical_models}, and briefly discuss the
recovery of the intrinsic moments in \S~\ref{sec:orbital_structure}.
We discuss our findings and conclude in
\S~\ref{sec:SRdiscussion_and_conclusions}.

\section{Galaxy models with a three-integral distribution function}
\label{sec:SRabelmodels}

Following \citetalias{2008MNRAS.385..614V}, we build triaxial Abel
models with a separable (or St\"ackel) potential that are a
generalization of the spherical Osipkov-Merritt
\citep[][]{1979PAZh....5...77O, 1985AJ.....90.1027M} models \citep[see
also][]{1991MNRAS.252..606D, 1999MNRAS.303..455M}.  These models, with
a DF $f(E,I_2,I_3)$ that depends on three exact integrals of motion,
allow for a large range of shapes and internal dynamics, while the
corresponding observables, including the LOSVD, can be computed
efficiently. They are well suited to simulate realistic imaging and
integral-field kinematics of early-type galaxies. Here, we construct
thirteen galaxy models, six nearly oblate axisymmetric, three nearly
prolate axisymmetric and four triaxial.  These galaxy models are
representative of early-type galaxies observed with the integral-field
spectrograph \Sauron\ \citep[e.g.][]{2004MNRAS.352..721E}, from oblate
fast-rotator to triaxial slow-rotator.

As in \S~4 of \citetalias{2008MNRAS.385..614V}, we use a triaxial isochrone
St\"ackel potential with a length scale of $10\arcs \simeq 1$\,kpc at the
adopted distance of $20$\,Mpc, and with a total mass of $10^{11}$\,\Msun.
Next, we choose the DF to be a linear combination of components of the form
$f(S) = (S-S_\mathrm{lim})/(1-S_\mathrm{lim})$, where both $S=-E+wI_2+uI_3$
and $S_\mathrm{lim}$ depend on the (constant) shape parameters $w$ and $u$.
Moreover, each DF component can be one of three types. The non-rotating (NR)
type is made of box orbits and tube orbits with both senses of rotation
populated equally. In addition we have two rotating types, long-axis rotating
(LR) and short-axis rotating (SR), which consist of tube orbits, and have net
rotation around the long axis and short axis respectively. We set $u$ and $w$
of each component such that the total density is similar to that corresponding
to the potential, except in the outer parts ($>30$\arcsec) where a steeper
surface brightness profile mimics the presence of dark matter (see also
\citetalias{2008MNRAS.385..614V}).

To calculate the observables we need to choose a viewing direction.
For the oblate/prolate axisymmetric models we adopt three
inclinations: close to edge-on/side-on $i=87$\dgr\ ($\cos i \simeq
0.05$), intermediate $i=60$\dgr\ ($\cos i = 0.50$), and close to
face-on/end-on $i=18$\dgr\ ($\cos i \simeq 0.95$). For the triaxial
models we use polar and azimuthal viewing angles of $\vartheta =
\varphi = 60$\dgr. We then compute for each DF component on a
rectangular grid on the sky plane the LOSVD and fit a Gauss-Hermite
series to obtain maps of the surface mass density $\Sigma$, and the
mean line-of-sight velocity $V$, velocity dispersion $\sigma$ and
higher-order Gauss-Hermite velocity moments up to $h_6$
\citep{1993MNRAS.265..213G,1993ApJ...407..525V}.  The surface
brightness is obtained from dividing $\Sigma$ by a constant (total)
mass-to-light ratio \ML$ = 4$\,\MLsun. To simulate \Sauron\ 
observations, we combine pixels to obtain a minimum signal-to-noise
(S/N), which we take proportional to the square root of the surface
brightness. Adopting a typical mean error of $7.5$\,\kms\ in $V$ and
$\sigma$, and $0.03$ for $h_3$ through $h_6$, we assign per (Voronoi)
bin an error, weighted with the S/N, and use these errors to
(Gaussian) randomize the kinematic maps.

The resulting maps of the surface brightness, $V$, $\sigma$, $h_3$ and
$h_4$ are shown in top rows in
Figures~\colfigref{fig:SRFR}--\colfigref{fig:SRST4} for each of the
thirteen galaxy models, for which the specific choices are as follows.

\emph{Fast rotator oblate models (FO1,FO2,FO3)} %
We choose the isochrone St\"ackel potential so that the (mean) axis
ratios of the corresponding density are $p_S = 0.99$ and $q_S = 0.76$,
such that the model is close to oblate axisymmetric\footnote{In the
  axisymmetric limit the isochrone potential reduces to the
  \cite{1962KK} potential \cite[cf.][]{1988ApJ...329..720D}.}. The DF
contains a NR and SR component with relative mass fractions of 30 and
70 per cent, and both with shape parameters $w=u=-0.5$. This results
in regular velocity field around the (photometric) minor axis. From a
luminosity weighted average of the simulated $V$ and $\sigma$ maps, we
calculate the (projected) specific angular momentum $\lambda_R =
\langle R |V| \rangle / \langle R \sqrt{V^2+\sigma^2} \rangle$
\citep[see][]{2007MNRAS.379..401E}. Both the edge-on and intermediate
inclination models FO1 and FO2, have $\lambda_R$ values of $0.36$ and
$0.33$ that are well within the regime of the fast rotators.  Even the
close to face-on model F03 has a $\lambda_R$ value of $0.14$ that is
above the boundary $\lambda_R=0.10$ between fast and slow rotators.

\emph{Slow rotator oblate models (SO1,SO2,SO3)} %
These three models are equivalent to the above fast rotator oblate models,
except that the relative mass fractions of the DF components are
inverted, i.e., 70 and 30 per cent for the NR and SR component,
respectively. As a result, they rotate significant slower, with
$\lambda_R$ values for the edge-on, intermediate and face-on
inclination models SO1, SO2 and SO3 of respectively $0.16$, $0.15$ and
$0.06$.

\emph{Slow rotator prolate models (SP1,SP2,SP3)} %
The isochrone potential in this case is chosen such that the axis
ratios of the corresponding density are $p_S = 0.79$ and $q_S = 0.78$,
i.e., close to prolate axisymmetric.  The DF consists of a NR and LR
component with relative mass fractions of 70 and 30 per cent, again
both with shape parameters $w=u=-0.5$. The resulting kinematics show
regular rotation around the (photometric) major axis. The amplitude of
rotation is rather mild, with $\lambda_R$ values for the side-on,
intermediate and end-on models SP1, SP2 and SP3 of respectively
$0.18$, $0.15$ and $0.05$.

\emph{Slow rotator triaxial models (ST1,ST2,ST3,ST4)} %
The axis ratios $p_S = 0.90$ and $q_S = 0.77$ of the density
corresponding to the chosen isochrone potential are consistent with a
triaxial model. Model ST1 only has a NR component with $w=u=-0.5$, and
hence shows no rotation ($\lambda_R = 0$), similar to \NGCN{4486}
(\textsc{M}\,87).  Model ST2 has an additional SR component with
relative mass fraction of 20 per cent, so that $\lambda_R = 0.14$.
Model ST3 has in addition to the NR component a SR component with
relative mass fraction of 10 per cent and different shape parameters
$w=0.5$ and $u=-1.0$. This results in a central compact kpc-size KDC
and zero rotation outwards due to the extended NR component, similar
to slow rotators like \NGCN{5831}. The $\lambda_R$ value within the
inner $10$\arcs\ rises above $0.10$, but at larger radii quickly drops
within the slow rotator regime.  Finally, model ST4 is the same as
model ST3, but with an additional LR component with $u=w=-0.5$. The
three NR, SR and LR components have relative mass fractions of 80, 10
and 10 per cent. Outside the KDC, the model has significant
(long-axis) rotation, similar to \NGCN{4365}. As a result, the
$\lambda_R$ value stays around $0.10$ in the outer parts.

Because the Abel models are constructed with a minimum number of DF
components, they are transparent and relatively quick to calculate,
but still they capture most of the rich dynamics observed in
early-type galaxies. Moreover, a careful decomposition of the observed
\Sauron\ two-dimensional stellar kinematics of early-type galaxies
using kinemetry shows that most early-type galaxies can be described
by only a few components \citep{2008MNRAS.390...93K}. Still, due to the
sometimes sharp transition between the DF components, the surface
brightness of the Abel models is not always as smooth as the surface
brightness of most early-type galaxies. As a result, the surface
brightness of the Abel models cannot accurately be fitted with
smooth parameterizations such as the Multi-Gaussian Expansion method
\citep[MGE;][]{1992A&A...253..366M, 1994A&A...285..723E},
unless negative Gaussians and/or strong twists are invoked. This would
not only be unnatural, but also incorrectly limit the possible
deprojections. Nevertheless, the MGE parameterization allows for a
straightforward deprojection and an efficient computation of the
gravitational potential that correspond to the resulting (triaxial)
intrinsic density. Therefore, we enforce a less accurate but smooth
MGE parameterizations of the surface brightness of the Abel models,
and show in \S~\ref{ssub:non_smooth_sb} that this does not
affect our main results.


\section{Constructing the Schwarzschild models} 
\label{sec:SRthe_modeling}

Based on the simulated observations of each of the thirteen Abel models, we
now construct triaxial \cite{1979ApJ...232..236S} models using our numerical
code described in detail in \citetalias{2008MNRAS.385..647V}. The code
reconstructs the intrinsic dynamical structure of a collisionless stellar
system with an arbitrary triaxial geometry by finding the superposition of
orbits that best fits simultaneously the (observed) photometry and stellar
kinematics in a self-consistent way. These models are significantly different
from axisymmetric Schwarzschild models \citep[e.g.][]{1998ApJ...493..613V,
2000AJ....119.1157G, 2004ApJ...602...66V, 2004MNRAS.353..391T,
2006MNRAS.366.1126C}, as they allow for triaxial intrinsic shapes and the
associated multiple distinct orbit families. Hence, they are applicable to
many of the early-type galaxies, including very boxy ones, and their triaxial
dark haloes. The models presented in this paper use the default parameters as
set in \citetalias{2008MNRAS.385..647V}, including the number of orbits and
the dimensions of the intrinsic polar mass grid. No regularisation was used
throughout this paper, to ensure that the results unbiased towards any form of
regularisation.

\subsection{Fitting the simulated observables of the Abel models} 
\label{ssub:fitting_abel_models}

The stellar kinematics fitted are the maps of the mean line-of-sight velocity
$V$, velocity dispersion $\sigma$ and higher-order Gauss-Hermite velocity
moments (up to $h_6$) computed for each of the Abel models. In addition, we
fit\footnote{Within the model the surface brightness and intrinsic density are
actually treated as constraints instead of being directly fitted.} the MGE
parameterization of surface brightness distribution and the corresponding
intrinsic luminosity density for a given viewing direction. Similarly, the
intrinsic mass density and gravitational potential is inferred from the
deprojected MGE parameterization of the surface mass distribution.
Because the surface mass distribution is not directly observable
(except perhaps in case of gravitational lensing), we constructed the Abel
models such the luminosity density mimics the mass density inside the observed
region (see \S~\ref{sec:SRabelmodels} and \citetalias{2008MNRAS.385..614V}).
Essentially, we implicitly assume that mass follows light. Unlike the
sometimes sharp transitions in the projected luminsity density as mentioned in
\S~\ref{sec:SRabelmodels}, the projected mass density of the Abel models is
always smooth and hence accurately parameterized by a MGE. The accelerations
in the (efficiently) computed MGE gravitational potential are accurate to $<
1\%$, when compared to the input analytic St\"ackel potential.

The types of orbits hosted by the gravitational potential depend strongly on
the intrinsic shape of the galaxy. This means that the capability of the
Schwarzschild models to fit the stellar kinematics (and the photometry in a
self-consistent way) also depends on the intrinsic shape (see for example
Fig.~5 of \citetalias{2008MNRAS.385..647V}). To obtain the intrinsic density
distribution from deprojection of the MGE parameterization three viewing
angles are needed that are typically unknown: the polar ($\vartheta$) and
azimuthal ($\varphi$) viewing angles, and the misalignment angle ($\psi$)
between the projected intrinsic short axis and the photometric minor axis.

Searching through a uniform grid in these three angles is unfeasible since the
deprojection is strongly non-linear. While in parts of the grid small changes
in the viewing angles can result in strong variations in the intrinsic shape,
in other regions the intrinsic shape remains nearly constant. Instead, we step
through the possible deprojections by regularly sampling three intrinsic shape
parameters: the axis ratios $p$ and $q$, and a third parameter $u$ that
represents the scale-length compression factor. This has the advantage that we
can sample the whole deprojection space with relatively few points and avoid
sampling in parts of the viewing space where no changes occur, while
increasing the sampling density where necessary (see also \S~3.7 of
\citetalias{2008MNRAS.385..647V}). For the tests in this paper it is
important to consider both the recovery of the viewing angles and the
intrinsic shape separately. As a good constraint on the shape does not
nessecarily mean a good constraint on the viewing angles and vice versa.

The ($p,q,u$) parameters are not equivalent to the St\"ackel $p_S$ and $q_S$
from \S~\ref{sec:SRabelmodels}: both describe the intrinsic shape, but the
latter describes the intrinsic shape measured directly from the moments of
inertia in the isochrone St\"ackel potential (see \S~4.1 in
\citetalias{2008MNRAS.385..614V}), whereas ($p,q,u$) is defined within the MGE
system.

The range of deprojections, and thus the sampling in the intrinsic
shape parameters, is different for each of the galaxy models presented
in this paper, as they have a different surface brightness. Typically
we have 125 deprojections in steps of 0.05 in $(p,q,u)$, and for each
deprojection we sample the mass-to-light ratio \ML\ at eleven
different values. In this way, we construct around 1400 Schwarzschild
models to the simulated observables of each Abel model.  Computing the
orbit library -- which has to be done for every deprojection -- takes
approximately six hours on a current single-core CPU. For each
deprojection and \ML\ the fitting takes another hour.  This means in
total about 2000 CPU hours to search the whole parameter space per
Abel model.

\subsection{Confidence criterion} 
\label{ssub:confidence_criterium}

To determine the best-fit model parameters and corresponding
uncertainties we compare for each Schwarzschild model its $\chi^2$
difference between the simulated observables and the corresponding
model predictions, weighted with the errors in the observables. The
(global) minimum in the $\chi^2$ yields the best estimate of the model
parameters. It is common practice, e.g., when dynamical models are
used to measure the mass of a black hole, to assign error bars based
on $\Delta\chi^2 = \chi^2 - \min{(\chi^2)}$ and the number of free
parameters $M$ of the model. In our case we have $M=4$ parameters
(three viewing angles and the mass-to-light ratio) after marginalizing
over the orbital weights. And thus our assiociated formal errors
become $\Delta\chi^2=4.72$ for 68\% confidence ($1\sigma$).

Whereas for the black hole mass measurements, with typically only a
few observations inside the sphere of influence of the black hole,
this works well, a different criterion is needed in our case of
measuring the intrinsic shape based on (simulated) integral-field
spectroscopy observations. The integral-field spectroscopy
observations that are used typically consist of several hundred (or
more) spectra. From each spectrum four (or more) velocity moments are
extracted, yielding a large number of individual constraints $N
\gtrsim 10^4$, which are fitted by the Schwarzschild model. Under
these circumstances the uncertainties in the $\chi^2$ itself become
important. The $\chi^2$ itself has a standard deviation of
$\sqrt{2(N-M)}$, which becomes non-neglibible in our case. Moreover,
we expect that systematic errors in the observations (e.g. due to
stellar template mismatch) can cause deviations in the $\chi^2$ of the
order of (or more than) the formal $1\sigma$.

Instead, we derive uncertainties on the best-fit parameters based on the
expected standard deviation of $\chi^2$, which we approximate to be
$\sqrt{2N}$ because $N \gg M$. Preliminary validation of this approach is
shown in \citetalias{2008MNRAS.385..647V}, where we construct triaxial
Schwarzschild models of one Abel model (similar to ST4) and the elliptical
galaxy \NGCN{4365} with an (apparent) kinematically decoupled core. In this
case and in the current study, the viewing angles of the Abel models are
known, so that this criterion for the standard error bars can be validated.

\section{Results of the dynamical models}
\label{sec:results_of_the_dynamical_models}

In Figures~\ref{fig:SRFR}--\ref{fig:SRST4}, we present for each Abel
model the simulated observables, the best-fit Schwarzschild model, and
the Schwarzschild model closest to the input parameters. The
difference between the latter two Schwarzschild models reveals the
(often small) differences due to the typical uncertainty allowed in
the fitted stellar kinematics (e.g.\ template mismatch errors). The
contour plots on the right indicate the confidence on the recovered
\ML, the shape and the corresponding viewing angles. Besides the
best-fit model (red crosses), they also show the location of the true
input values (green diamonds). The white areas do not have a feasible
deprojection. We first discuss the individual cases.

\def \Srfifh {6.0cm}

\begin{figure*}
\centering
FO1 \\
\includegraphics[height=\Srfifh]{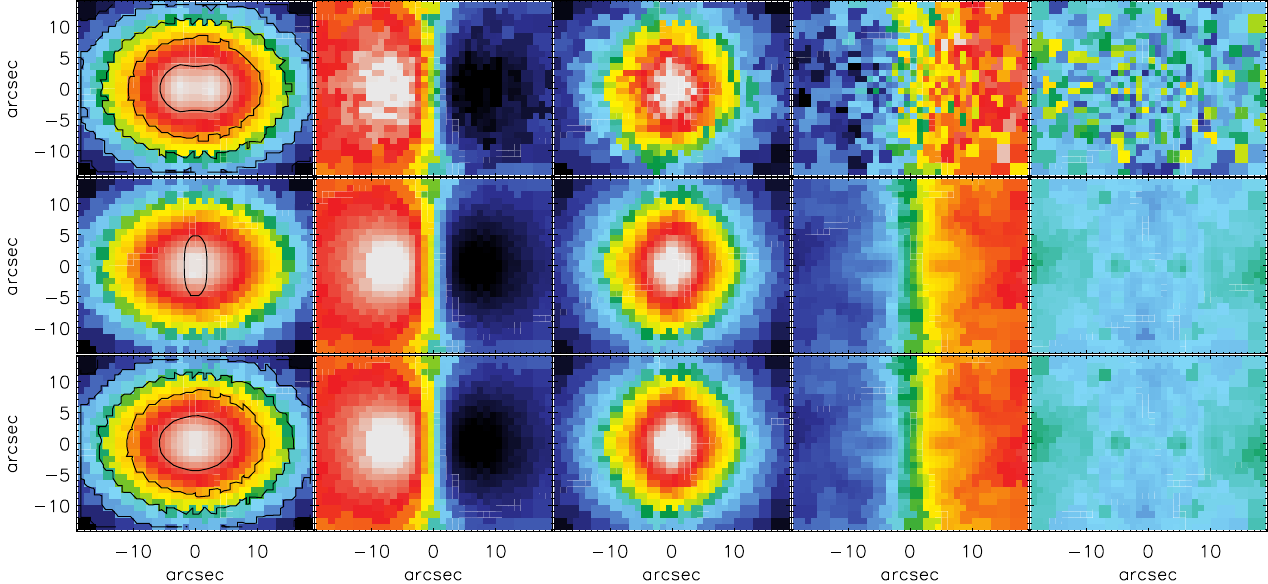}
\includegraphics[height=\Srfifh]{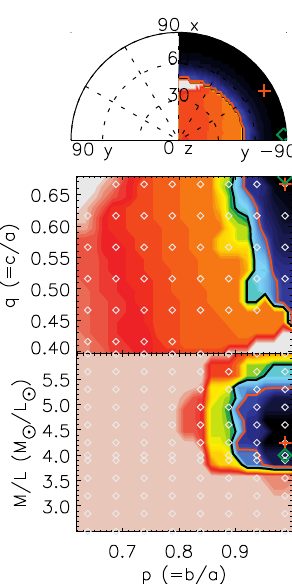}
\\ FO2\\
\includegraphics[height=\Srfifh]{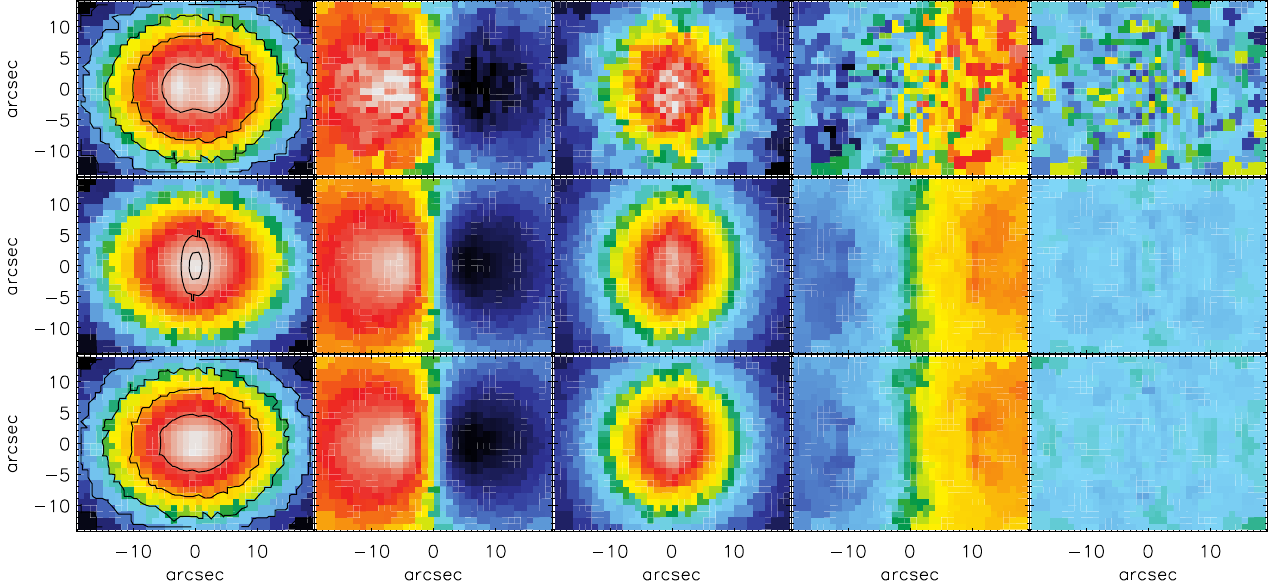} 
\includegraphics[height=\Srfifh]{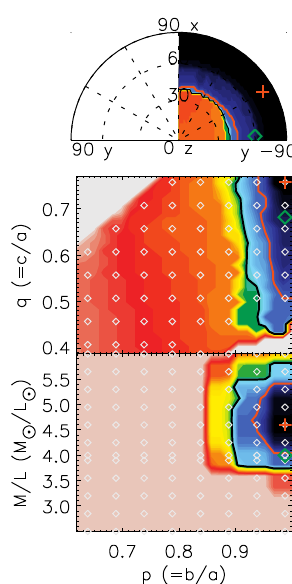} 
\\ FO3 \\
\includegraphics[height=\Srfifh]{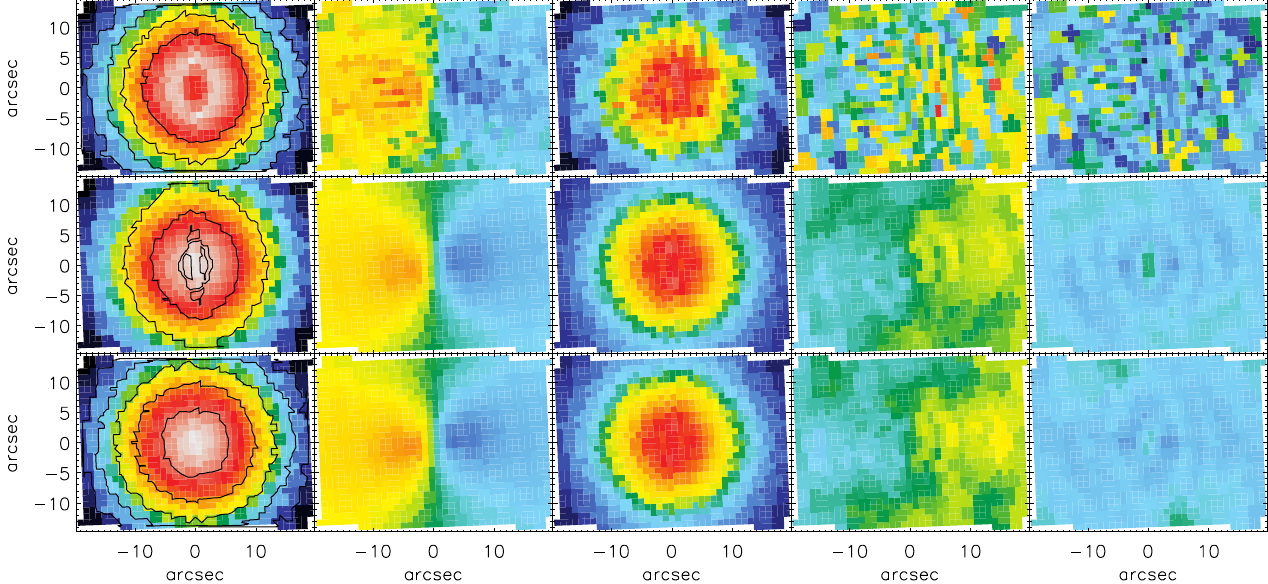} 
\includegraphics[height=\Srfifh]{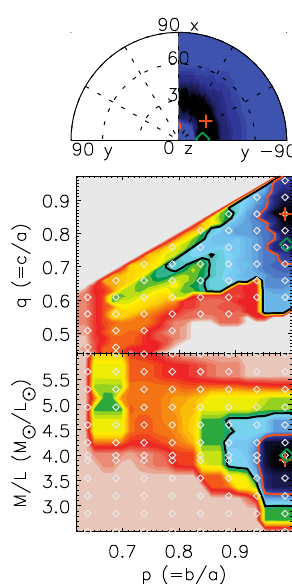} 

\caption{Dynamical modelling of fast rotator oblate models FO1, FO2
  and FO3. Each plot shows the stellar kinematics of the Abel model
  (top), the best-fit Schwarschild model (middle), and the fit at the
  input parameters (bottom). The contours on the surface brightness
  image (leftmost panel) show the observed surface brightness (top),
  the surface brightness of the MGE model (bottom) and the difference
  between the two (middle). Colums 2 through 5 show the line-of-sight
  velocity moments: $V$ (-110\ldots110 \kms), $\sigma$ (100\ldots260
  \kms), $h_3$ and $h_4$ (-0.1\ldots0.1). The panels on the right side
  show marginalized confidence regions of the viewing angles
  $\vartheta$ and $\varphi$ (in Lambert projection seen down the
  $z$-axis, so that $\vartheta$ runs radially and $\varphi$ runs
  azimuthally), $p$ versus $q$, and $p$ versus \ML. The inner red and
  outer black contour signify 1$\sigma$ and 3$\sigma$ confidence,
  repectively. The red cross and green diamond highlight the best-fit
  parameters and the original input parameters, respectively. }

\label{fig:SRFR}
\end{figure*}             

\def \Srfifh {6.58cm}

\begin{figure*}
\centering
SO1 \\
\includegraphics[height=\Srfifh]{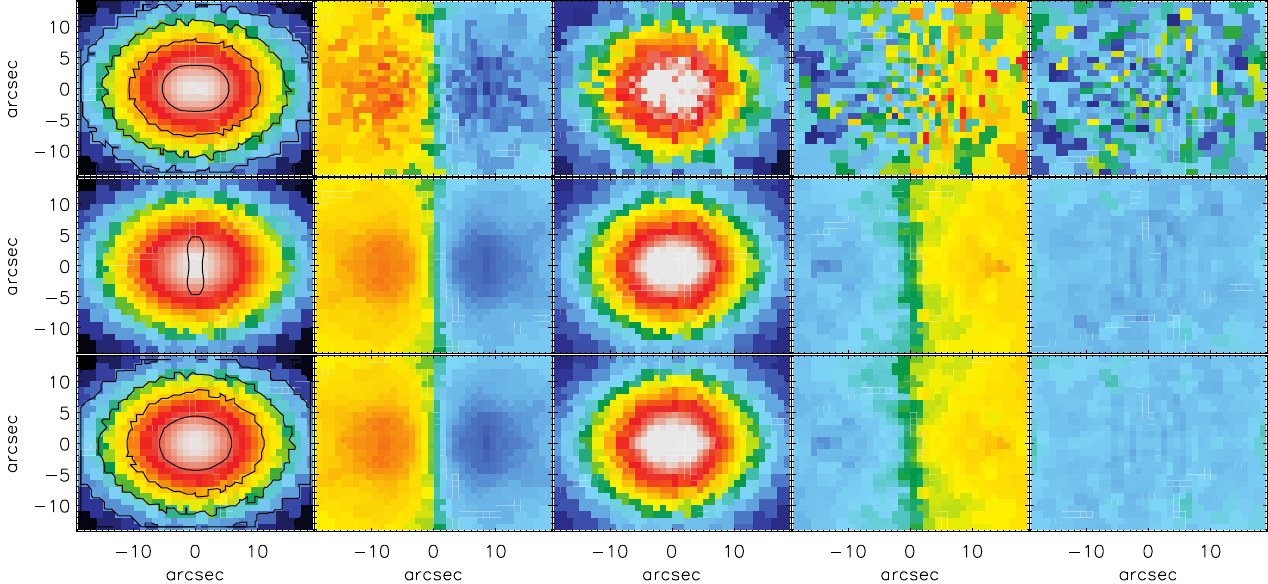}
\includegraphics[height=\Srfifh]{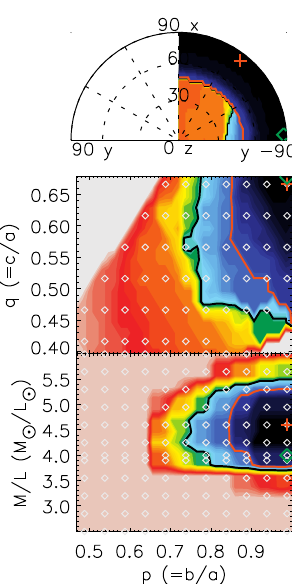}
\\ SO2 \\ \includegraphics[height=\Srfifh]{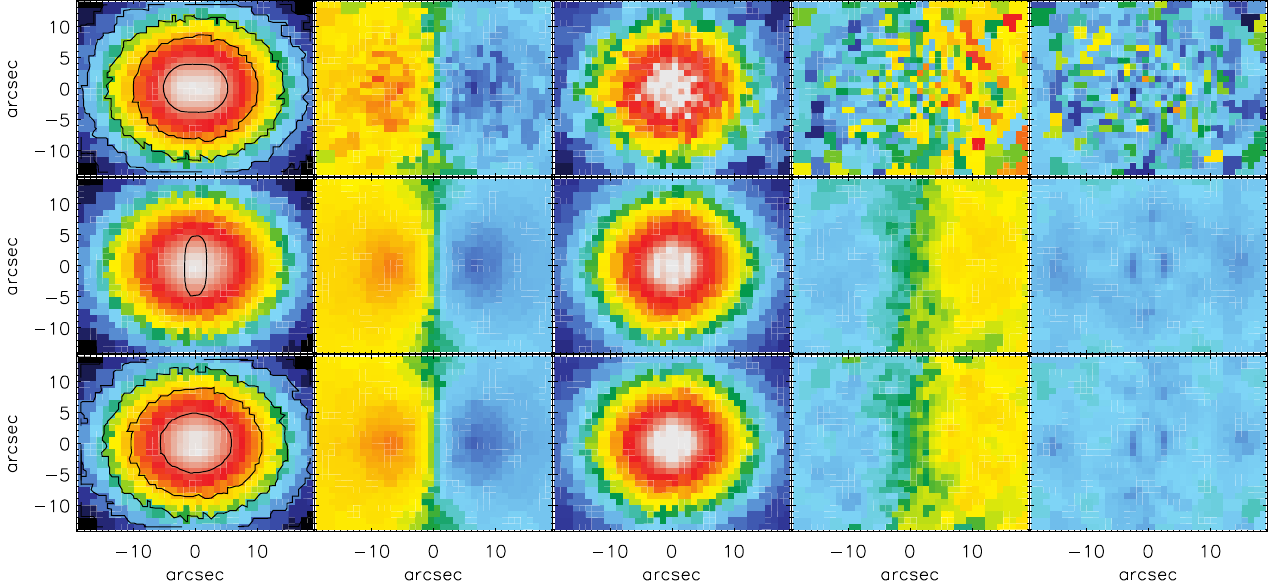} 
\includegraphics[height=\Srfifh]{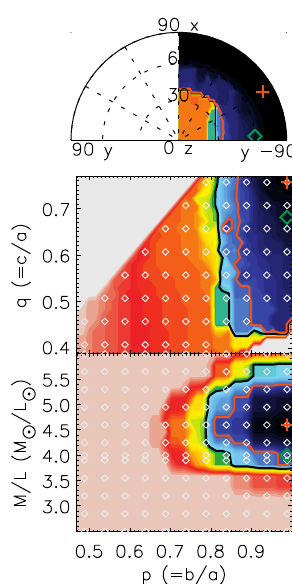} 
\\ SO3 \\
\includegraphics[height=\Srfifh]{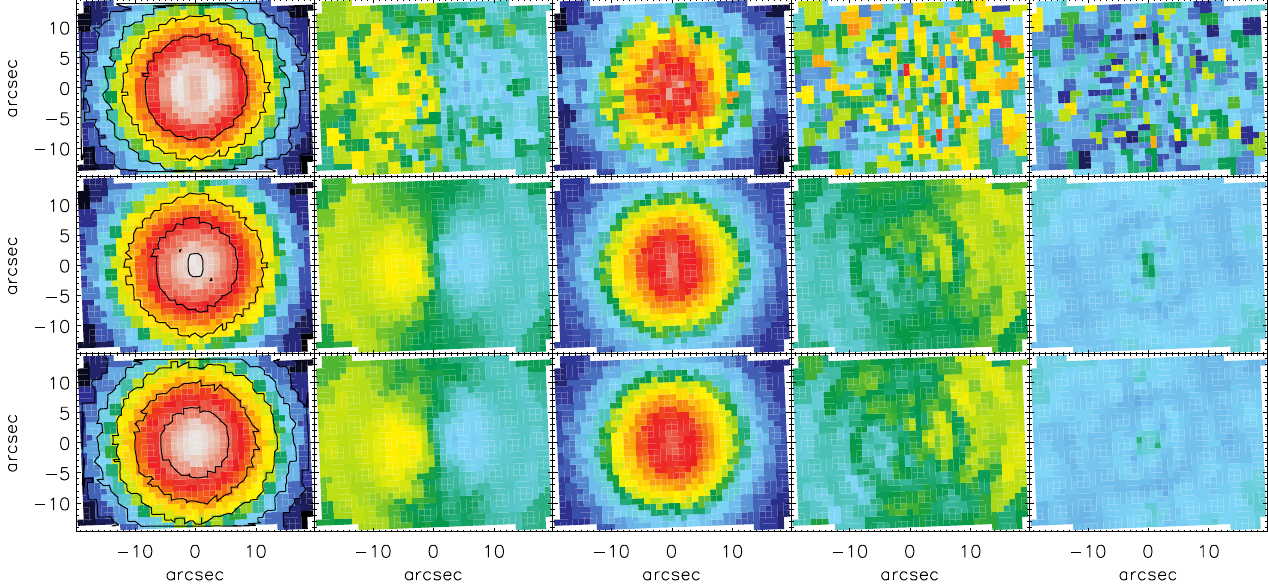} 
\includegraphics[height=\Srfifh]{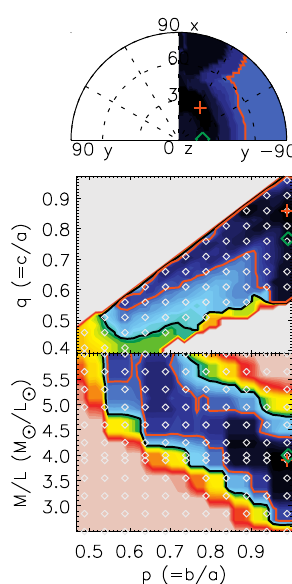} 
\caption{Slow rotator oblate models SO1, SO2 and SO3. See Fig.~\ref{fig:SRFR} for description.}
\label{fig:SROR}
\end{figure*}             


\begin{figure*}
  \centering
  \textsc{SP1 }\\

  \includegraphics[height=\Srfifh]{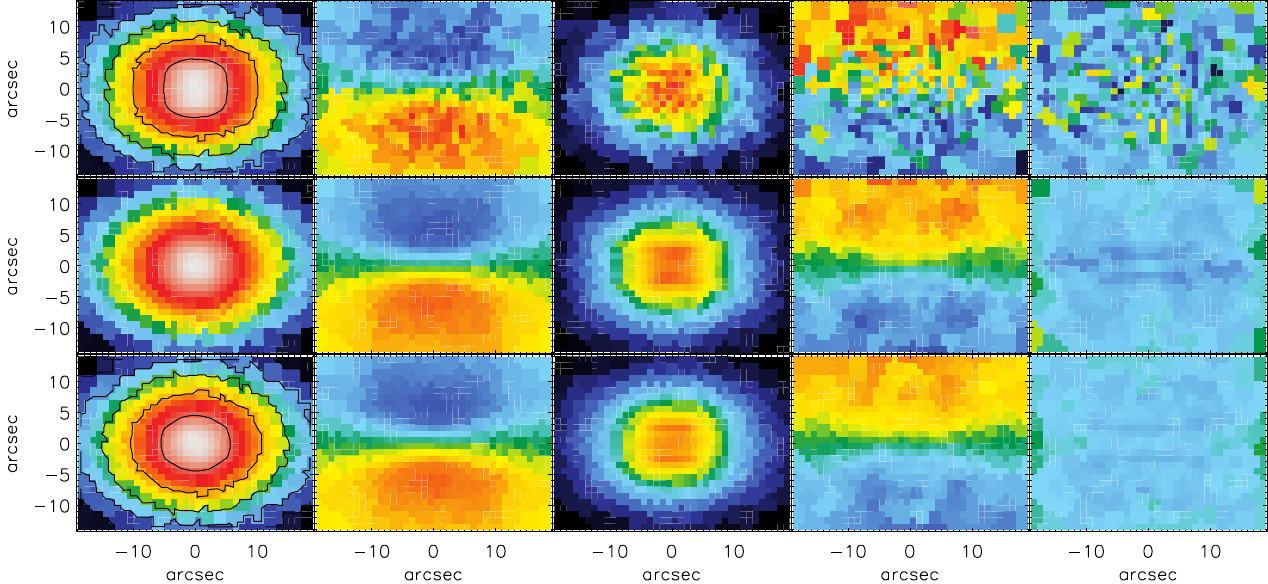}
  \includegraphics[height=\Srfifh]{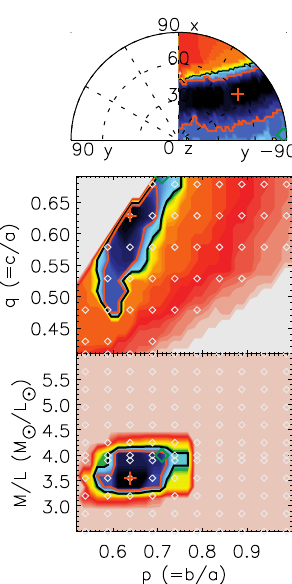}
  \\\textsc{ SP2} \\

  \includegraphics[height=\Srfifh]{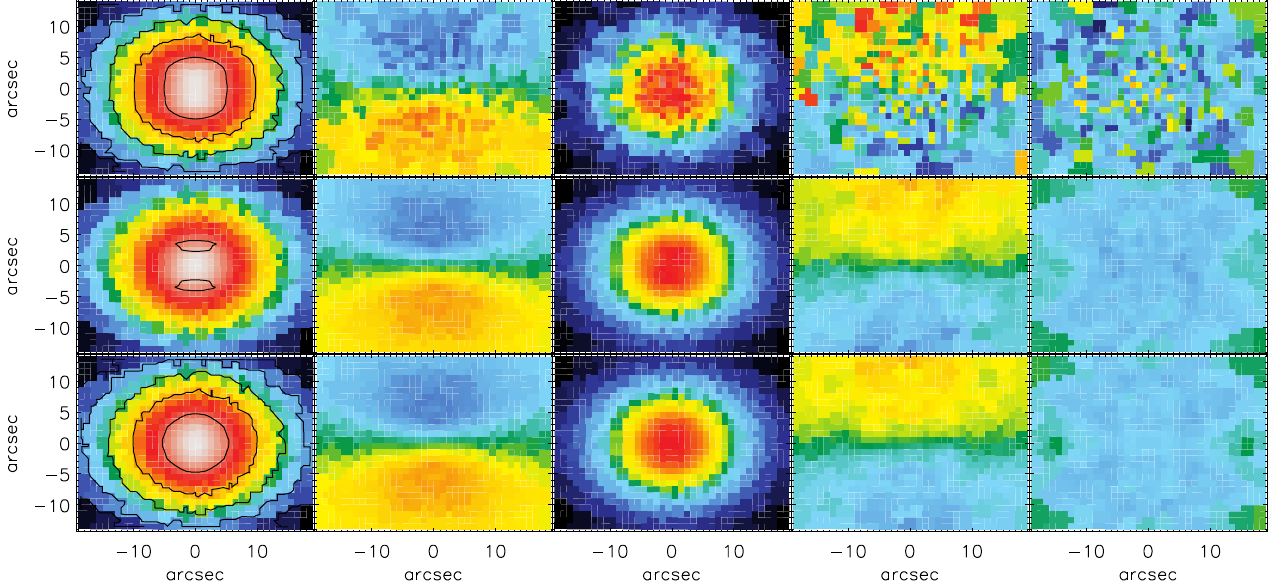}
  \includegraphics[height=\Srfifh]{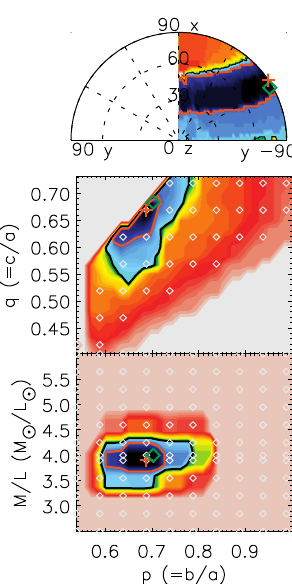}
  \\ \textsc{SP3} \\

  \includegraphics[height=\Srfifh]{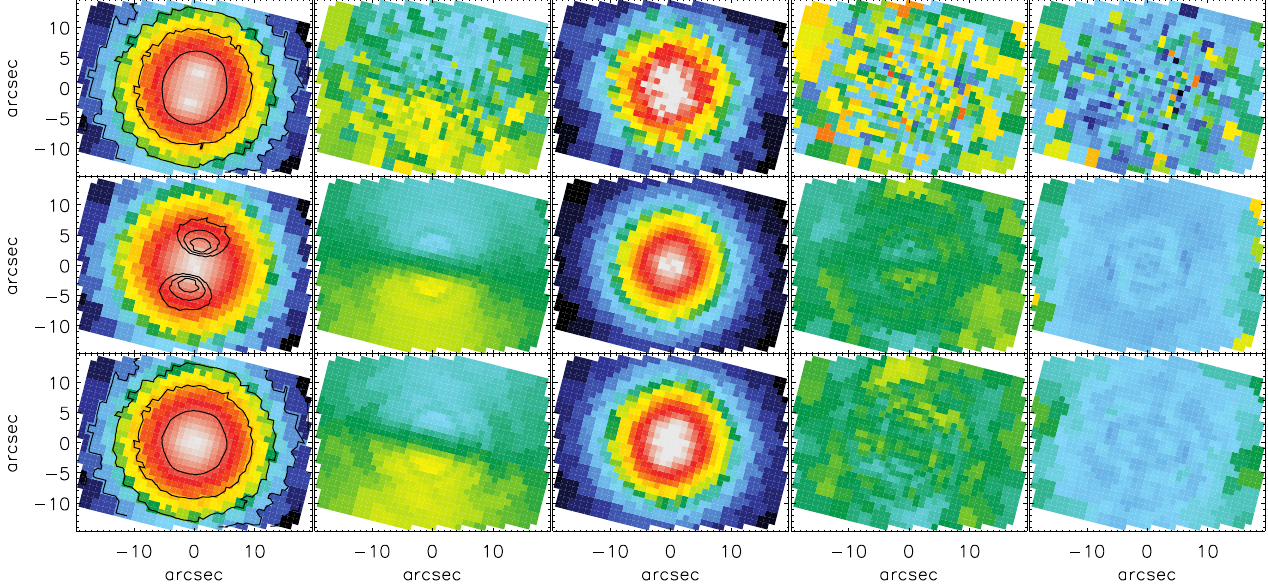}
  \includegraphics[height=\Srfifh]{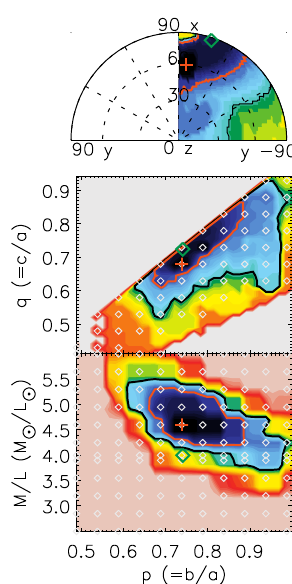}
\caption{Slow rotator prolate models SP1, SP2 and SP3. See Fig.~\ref{fig:SRFR} for description.}
\label{fig:SRSP}
\end{figure*}

\begin{figure*}
  \centering
  \textsc{ST1} \\

  \includegraphics[height=\Srfifh]{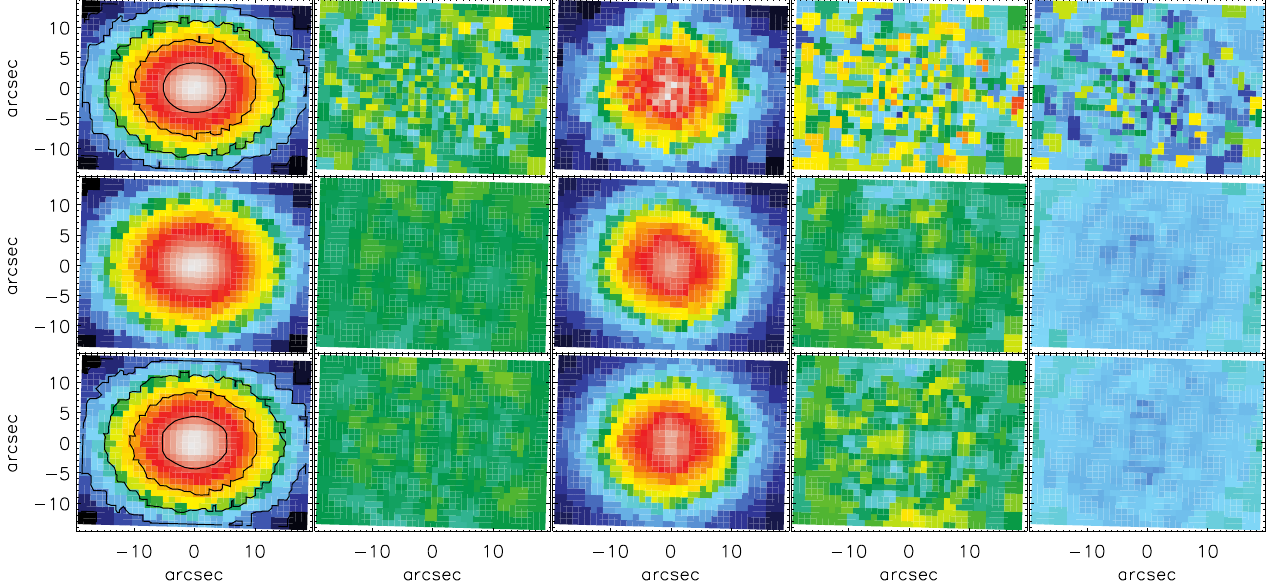}
  \includegraphics[height=\Srfifh]{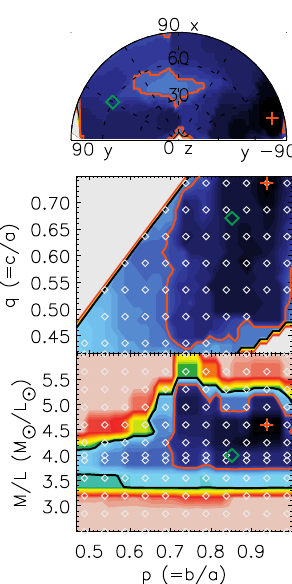}
  \\ ST2 \\

  \includegraphics[height=\Srfifh]{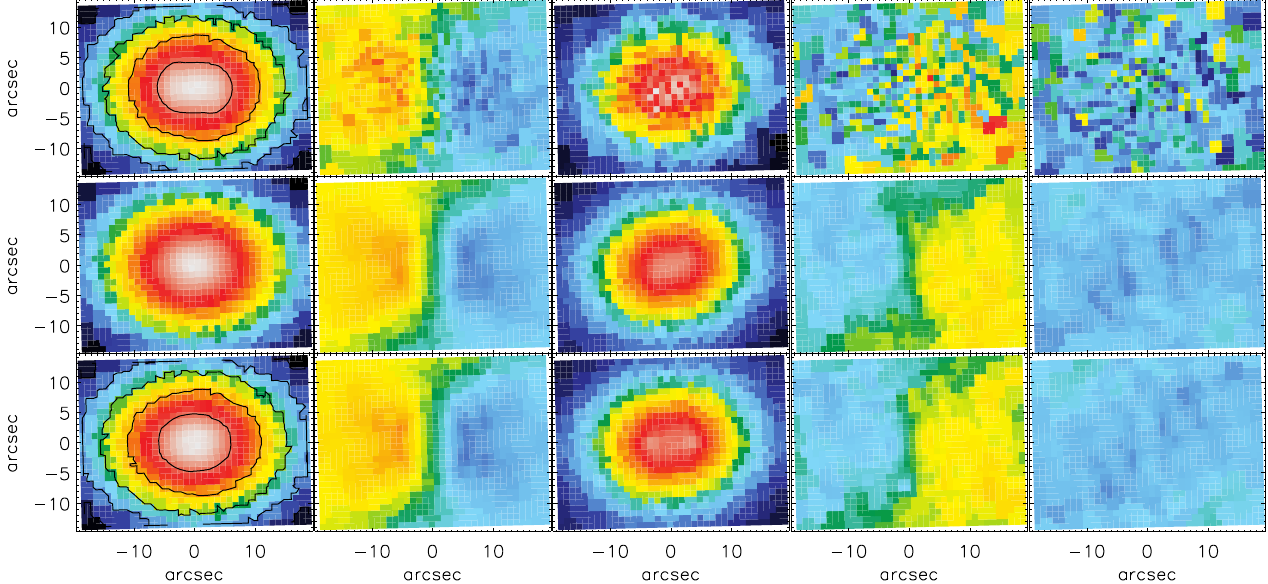}
  \includegraphics[height=\Srfifh]{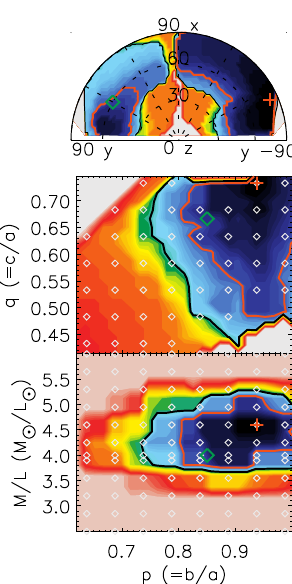}
  \\ ST3 \\

  \includegraphics[height=\Srfifh]{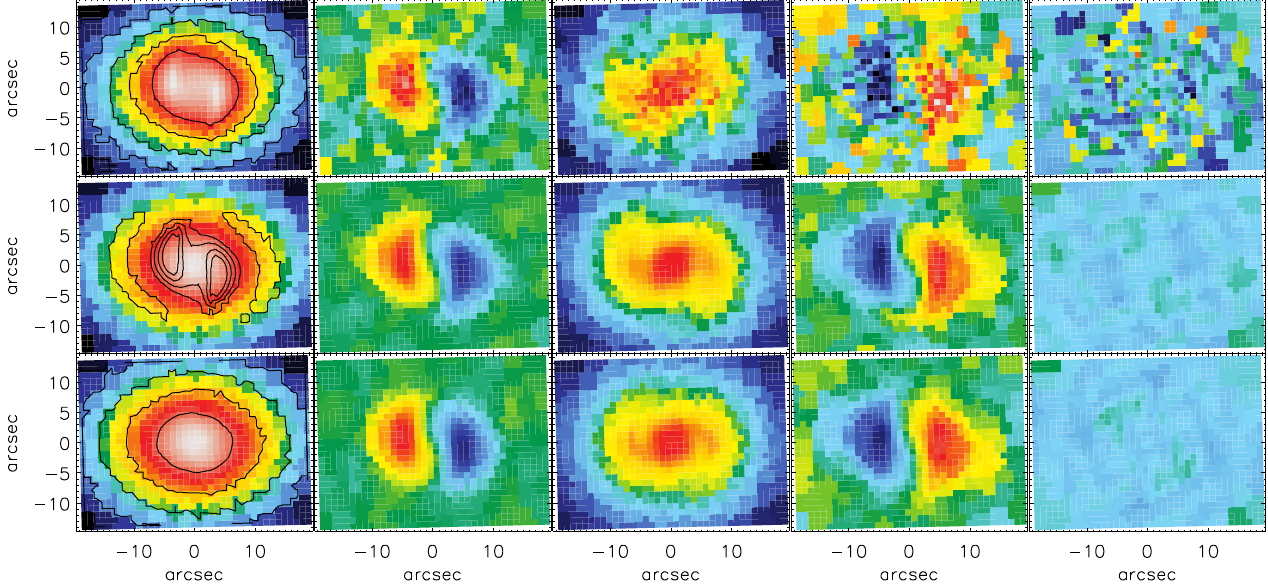}
  \includegraphics[height=\Srfifh]{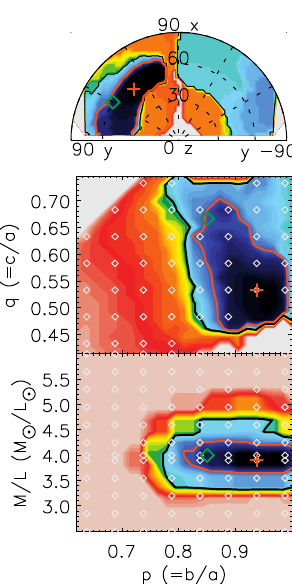}
\caption{Slow rotator triaxial models ST1, ST2 and ST3. See Fig.~\ref{fig:SRFR} for description.}
\label{fig:SRST}
\end{figure*} 

\begin{figure*}
  \centering
  ST4 \\

  \includegraphics[height=\Srfifh]{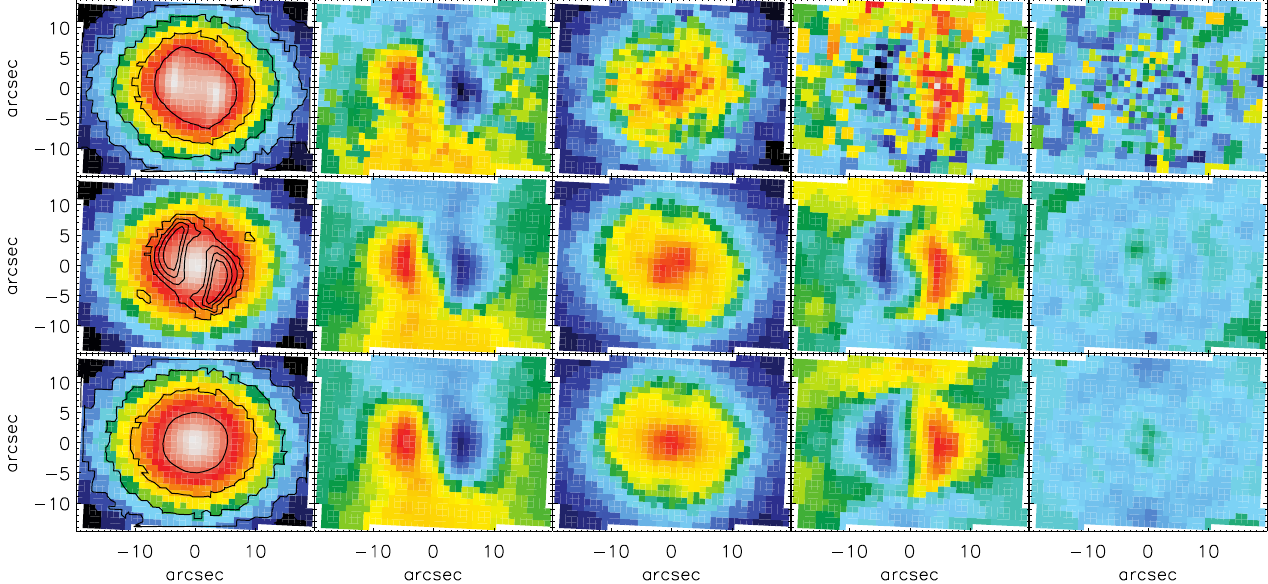}
  \includegraphics[height=\Srfifh]{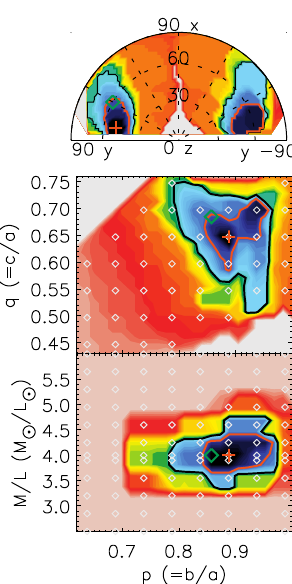}
\caption{Slow rotator triaxial model ST4. See Fig.~\ref{fig:SRFR} for description.}
\label{fig:SRST4}
\end{figure*}

\subsection{The oblate models}
\label{ssub:soblatemods}

For all six oblate Abel models (SO1 through FO3) shown in
Figures~\colfigref{fig:SRFR} and~\colfigref{fig:SROR}, the
Schwarzschild models around the minimum in $\chi^2$ are (very) good
fits with reduced $\chi^2$ per degrees-of-freedom close to unity. The
shape is recovered well within the $1\sigma$ confidence region. The
edge-on and intermediate models (SO1, SO2, FO1, FO2) have a strong
constraint on $p>0.95$, fully excluding strongly triaxial geometries,
while $q$ is generally unconstrained, apart from the lower limit $q >
1-\eps$ implied by the largest value of the observed ellipticity
$\eps$.  Hence, we cannot constrain the inclination of these (near)
axisymmetric systems. The differences in the fitted kinematics between
the best-fitting model and the model with the correct input parameters
(see Fig.~\ref{fig:SRFR}) are so small that they are negliable
compared to possible (systematic) errors in real observations.

As expected, the face-on slow-rotator model SO3 has the worst recovery with no
significant constraint on its shape and a wide range in \ML\ allowed. All the
other oblate models have a constraint on \ML\ that is consistent with the
input value. The constraints on all the fast-rotator oblate models are tighter
than those on the slow-rotator oblate models because the mean velocity
amplitude in the fast-rotator models is larger, which effectively shrinks the
simulated observational errors.

For all oblate models the flattening $q$ of the best-fit model is never
far removed from the input value $q_0$: $\Delta q \equiv |q - q_0| <
0.08$. Even though the viewing direction, or the inclination in this
case, is mostly unconstrained, the latter indicates that the intrinsic
shape is not fully degenerate. Observations that are more accurate
and/or span a larger field-of-view are expected to yield tighter
constraints on the intrinsic shape. When we fitted the surface
brightness of the Abel models directly instead of the overly smooth MGE
parameterization, the constraints on the shape improve, but only
mildly, showing that our MGE parameterization is sufficient for the
oblate models.

\subsection{The prolate models}
\label{ssub:the_prolate_models}

The results for the (slow-rotator) prolate models (SP1, SP2, SP3) are
shown in Fig.~\colfigref{fig:SRSP}. The constraint on the shape is
better than for the oblate models and the true shapes and \ML\ lie
within or on the 1$\sigma$ confidence contour. In all cases the axis
ratios are well recovered with allowed ranges $\Delta p < 0.07$ and
$\Delta q < 0.07$. Even the end-on model SP3 has a good
  constraint on the shape, excluding more than 75\% of the allowed
  range in $p$ and $q$. The constraint on the shape for prolate
  systems is better than for the oblate systems due to the fact that
  there are less (near) prolate deprojections possible than oblate
  ones.  Similarly, the limits on the inclination from the photometry
  are more restrictive in the prolate than oblate case, but still the
  regular kinematics do not really further constrain the inclination.

\subsection{The triaxial models}
\label{ssub:the_triaxial_models}

Model ST1 (Fig.~\colfigref{fig:SRFR}) is a non-rotating round galaxy
seen at an intermediate viewing angle. Like the face-on slow-rotator
oblate model SO3 - which shows almost no rotation - the triaxial
Schwarzschild models are capable of reproducing all features in the
simulated observables at nearly all viewing angles. The \ML\ is
recovered with a similar accuracy as in the case of the oblate models,
despite the lack of constraint on the intrinsic shape. Even so, the
best-fit $p$ and $q$ are with 10 per cent from the input values,
while the formally allowed ranges are larger with $\Delta p \sim 0.30$
and $\Delta q \sim 0.30$.

The zero velocity curve of the stellar kinematics in model ST2 seen in
Fig.~\colfigref{fig:SRST} is twisted with respect to the photometric
minor axis. Since this is only possible in a non-axisymmetric system,
we might expect a rather good recovery of the intrinsic shape.
However, the constraints are not very good with ranges $\Delta p \sim
0.15$ and $\Delta q \sim 0.25$, and comparable with those on the shape
of the slow rotator oblate models SO1 and SO2. Model ST3
(Fig.~\colfigref{fig:SRST}) has a central rotating component and no
rotation in the outer parts. The resulting constraint on the intrinsic
shape with ranges $\Delta p \sim 0.15$ and $\Delta q \sim 0.20$ is
only slightly better than model ST2.  However, as we show below in
\S~\ref{ssub:non_smooth_sb} and in Fig.~\ref{fig:Striax2alt}, taking
into account the strong photometric twist improves the recovery of the
intrinsic shape as well as the (weak) constraint on the viewing
angles.

Model ST4 (Fig.~\colfigref{fig:SRST4}) contains a central KDC like
model ST3, and additionally shows significant rotation in the outer
parts. Although the inner part of the surface brightness is not well
reproduced by the smooth MGE parameterization, unlike model ST3, the
intrinsic shape is nevertheless accurately recovered and constrained,
with ranges $\Delta p < 0.05$ and $\Delta q < 0.05$. The triaxial Abel
model used in \citetalias{2008MNRAS.385..647V}\ and
\citetalias{2008MNRAS.385..614V} is similar to model ST4, but at a
different viewing direction and with more photometric twist in the MGE
parameterization, resulting in (slightly) tighter constraints on the
intrinsic shape as well as on the viewing angles.

\subsection{Fitting the non-smooth surface brightness directly}
\label{ssub:non_smooth_sb}

\begin{figure}
\centering 
\hspace{0.5cm} ST2 \hspace{3cm} ST3\\
\includegraphics[height=3in]{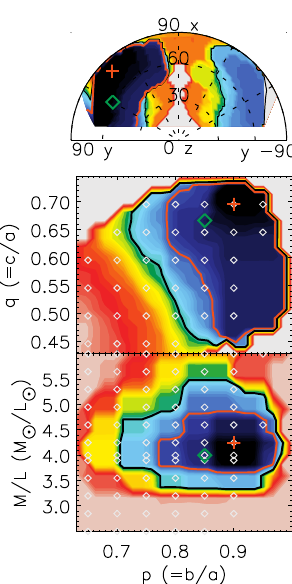}
\includegraphics[height=3in]{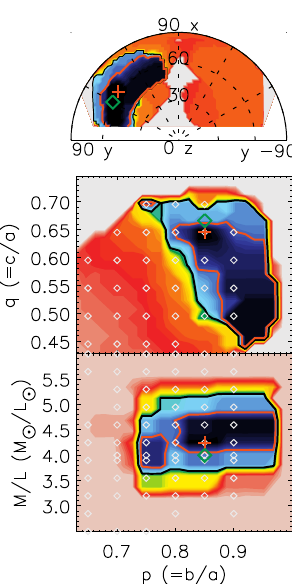}    
\caption{The shape recovery of model ST2 (left) and ST3 (right), when
  the observed surface brightness is fitted directly, bypassing the
  (too) smooth MGE parameterization. The presentation of the models
  here is identical to the ones in in Fig.~\colfigref{fig:SRST}.  }
\label{fig:Striax2alt}
\end{figure}

As mentioned in \S~\ref{sec:SRabelmodels}, the surface brightness of
the Abel models can show a sudden transition between DF components,
which cannot be accurately described by a smooth MGE parameterization
without invoking negative Gaussians and/or strong photometric twists.
A way to avoid this is to bypass the MGE and use the observed surface
brightness directly. This means that the intrinsic density is not
constrained anymore, since it follows from deprojecting the MGE
parameterization. In Fig.~\ref{fig:Striax2alt}, we present the results
for model ST3, which shows most evidently both a central depression as
well as a strong photometric twist in its surface brightness. The
intrinsic shape is still correctly recovered, but now the best-fit
model is much closer to the input values. Also many of the
deprojections with $p>0.93$ or $q>0.67$ are excluded, leading to a
significantly tighter constraint on the intrinsic shape then when the
MGE parameterization is used. In Fig.~\ref{fig:Striax2alt} we also
show the same for ST2, for which the effect is less pronounced but
still significant: the best-fit model is now much closer with $p$ and
$q$ less than $0.05$ away from their input values, but the confidence
intervals are unchanged.

This shows that the MGE representation we used in
\S~\ref{ssub:the_triaxial_models} is too smooth, adding irrelevant
deprojections that (artificially) increase the error bars on $p$ and
$q$. However, there is a delicate balance between reproducing too
little and too much photometric twist. Fitting unrealistic sharp
features (due to for example dust in real observations), may exclude
allowed deprojections or even the true underlying intrinsic shape. To
avoid the latter, the fit to the surface brightness should be as round
as possible, while still accurately describing the observed
photometry. 

\section{Intrinsic velocity moments} 
\label{sec:orbital_structure}

\begin{figure*}
\centering 
\includegraphics[height=8cm,clip,trim=0.cm 0cm 0.57cm 0cm]{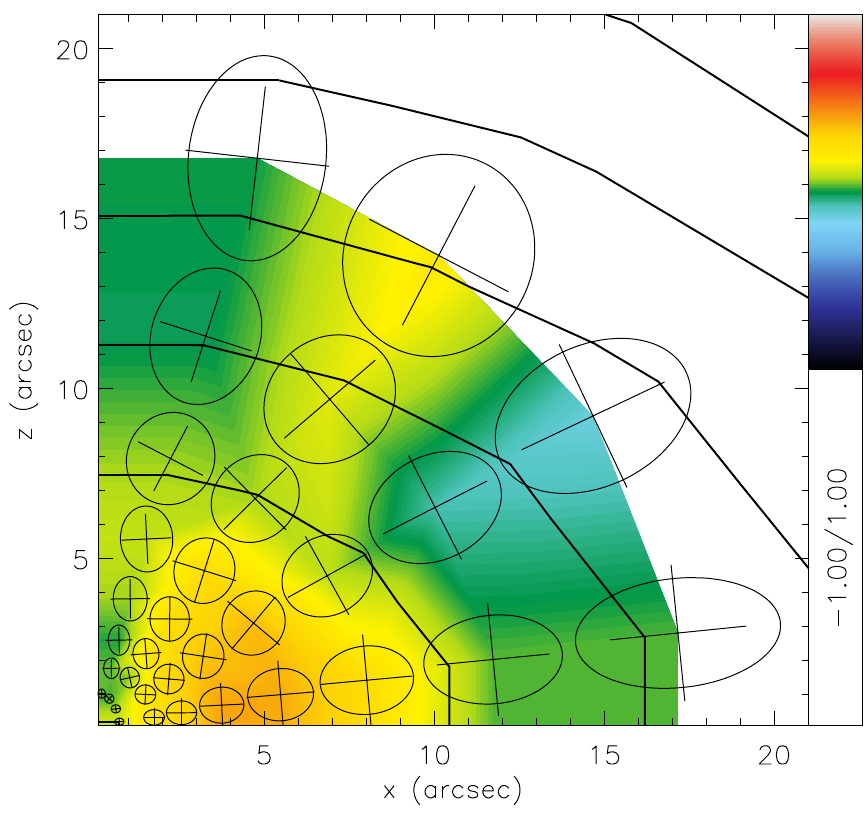}
\includegraphics[height=8cm,clip, trim=1.0cm 0cm 0.0cm 0cm]{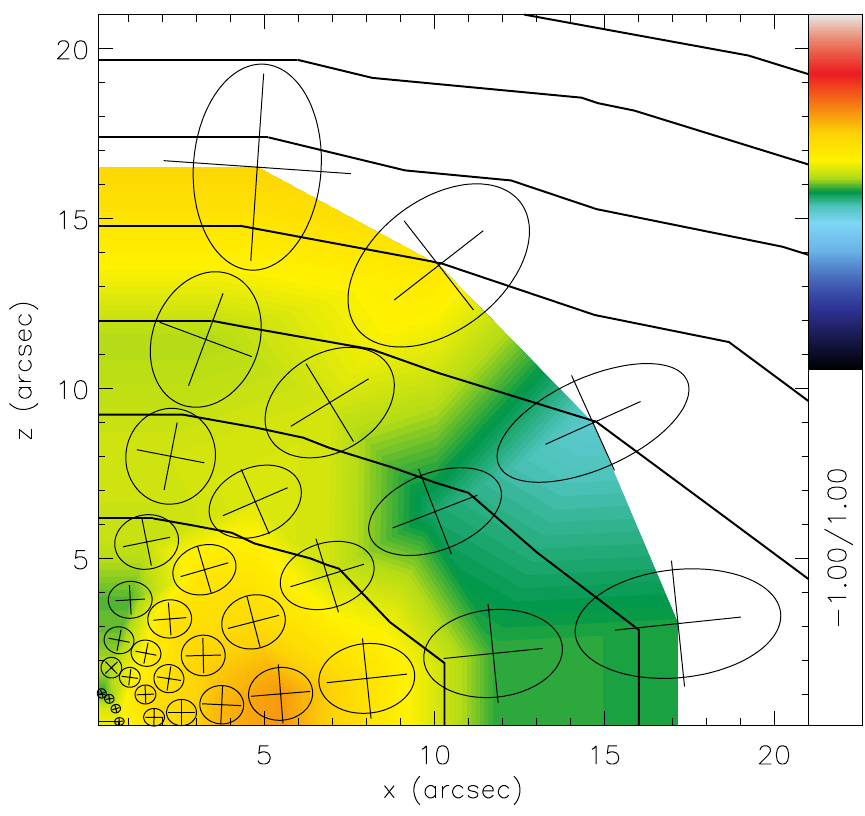} 
\caption{The intrinsic velocity moments of
  triaxial Schwarzschild models fitted to the simulated observables of
  the Abel model ST3. In the left panel, the intrinsic shape
  parameters are at the true input values, while in the right panel
  they are at the boundary of the corresponding $1\sigma$ confidence
  region. The colour represents the (intrinsic) $V/\sigma$, and the
  ellipses are cross-sections of the velocity ellipsoid with the
  $(x,z)$ plane, while the crosses indicate the (relative) size of the
  velocity ellipsoid in the perpendicular ($y$-axis) direction (see
  \S~5.3 of \citetalias{2008MNRAS.385..614V} for further details). The
  black curves are contours of constant mass density in steps of 1
  mag. \label{fig:SRvsig}}
\end{figure*}

In \citetalias{2008MNRAS.385..614V}, we have shown that given the
correct input viewing direction, the internal dynamics and
three-integral distribution function can be recovered. Here, we
investigate what happens to the intrinsic velocity moments when the
viewing angles are allowed to deviate from the input values. We expect
the intrinsic velocity moments, in terms of $V/\sigma$ and velocity
anisotropy, to vary when the viewing direction and hence intrinsic
shape changes.  Still, if we stay within the $1\sigma$ confidence
regions of the fitted intrinsic shape parameters, these variations
might be limited due to the constraints coming from the
two-dimensional kinematics. We use model ST3 from the previous section
as an illustration. In Fig.~\ref{fig:SRvsig}, we show the intrinsic
velocity moments at the input values $(p,q)=(0.85,0.66)$ and at
$(p,q)=(0.90,0.50)$, which is at the (lower) boundary of the
corresponding $1\sigma$ confidence region (see also
Fig.~\ref{fig:Striax2alt}). While the mass density of the two models
(indicated by the black contours) is clearly different, the intrinsic
velocity moments are quite similar. This suggests that as long as the
intrinsic shape is reasonably well constrained, the internal dynamics
are also reliably recovered.

\section{Discussion and conclusions}
\label{sec:SRdiscussion_and_conclusions} 

We constructed thirteen galaxy models with simulated photometry and
two-dimensional stellar kinematics that represent the \Sauron\
observations of early-type galaxies, from oblate fast rotator to
triaxial slow rotator. We fitted realistic three-integral triaxial
Schwarzschild models to the simulated observables of these Abel models
and we used the $\chi^2$ difference to investigate how well the
intrinsic shape as well as the mass-to-light ratio can be recovered.
Due to the large number of observational constraints (typically $N >
10^4$) with possible systematic errors (e.g.\ stellar template
mismatch) in the case of real integral-field observations, the
commonly used $\Delta \chi^2 = \chi^2 - \min{(\chi^2)}$ is not useful
to estimate the uncertainties on the model parameters. Instead, we
showed that the standard deviation $\sqrt{2N}$ works well as a
confidence criterion.

The simulated observations of the (near) oblate Abel models that are not
viewed close to face-on strongly excluded triaxial intrinsic shapes with
$1-p<0.05$. The flattening $q$ is almost only constrained by the photometry,
although the models with more projected rotation provide a slightly tighter
constraint since the relative errors in the odd velocity moments are smaller.
This means that for fast-rotator early-type galaxies, which seem to be
consistent with oblate axisymmetry, the inclination is not significantly
further constrained by current two-dimensional stellar kinematics, consistent
with \cite{2005MNRAS.357.1113K}. Similarly, for the (near) prolate Abel models
that are not viewed close to end-on a triaxial intrinsic shape is strongly
excluded with $p-q<0.05$, but the viewing direction is only weakly
constrained.

Unlike axisymmetric galaxies, a triaxial galaxy can show twists in
both the surface brightness and velocity field (and in higher-order
velocity moments), as well as misalignment between the kinematic
(rotation) axis and the photometric minor axis. The stronger such
features are in the observations, the more accurate the intrinsic
triaxial shape can be recovered, often better than in the axisymmetric
case. In particular, for the triaxial Abel models with a central KDC,
$p$ and $q$ can be recovered within 5\%, especially when (misaligned)
rotation is present in the outer parts. Since such kpc-size KDCs seem
typical for the slow-rotator early-type galaxies, we expect that their
intrinsic shape can be well estimated from dynamical modeling.
However, when inferring the mass model from a fit to the surface
brightness, the fit should be as round as possible to not exclude
allowed deprojections, but still accurately describe the photometry,
including twists, to avoid artificially increasing the range of
allowed intrinsic shapes.

Near face-on oblate galaxies and end-on prolate galaxies appear round
on the sky and show almost no (projected) rotation. We showed that in
these cases, as well as for triaxial galaxies with no intrinsic (net)
rotation, the best-fit $p$ and $q$ are still remarkably close to their
true values (typically within 10\%), but the formal uncertainties are
much larger (typically around 30\%).  This means that for galaxies
that appear round on the sky and show little rotation, the only hope
of recovering the intrinsic shape and corresponding viewing direction
is through additional constraints such as provided by a thin embedded
disc (in either dust, gas or stars).

In all cases, the \ML\ is correctly recovered, within about 10\% when
the intrinsic shape is well recovered, increasing up to 20\% for the
models with a poor constraint on the intrinsic shape. This is in line
with \cite{2007MNRAS.381.1672T}, who showed that applying axisymmetric
Schwarzschild models to triaxial merger remnants results in a larger
scatter in the recovered \ML\ compared to oblate remnants. However, as
long as generic dynamical models such as our triaxial Schwarzschild
models are used, the recovery of the \ML\ is nearly independent of the
underlying intrinsic shape. This is not unexpected given the good
correlation between dynamical and virial \ML\ estimates
\citep[e.g.][]{2006MNRAS.366.1126C}, and the observed tight scaling
relations in early-type galaxies such as the fundamental plane
\citep[e.g.][]{Dressler1987,Djorgovski1987}.

In real galaxies the photometry and the kinematics are usually
measured independently and as a result their relative alignment is not
perfectly known. This can create a small kinematic misalignment in an
otherwise perfectly axisymmetric galaxy, or vice-versa. To test the
significance of this we re-ran the fast rotator oblate model FO2 with
a positive and negative misalignment of 2\dgr, emulating a typical
measurement error on the PA (in a fairly round galaxy). The resulting
recovered parameters and their errors were very similar to the
original FO2 model, with the $\chi^2$ only marginally higher and the
$3\sigma$ confidence region slightly extended. This shows that the
uncertainty in the relative PA in these observations does not effect
the recovery of the intrinsic shape nor the recovery of the
mass-to-light ratio. 

Furthermore, we ran seperate models of FO2 and ST4 in which we doubled the
number of orbits. The resulting models, including the $\chi^2$ contours and
the recovered kinematics, where nearly identical showing that the number of
orbits used in our models is sufficient.

We conclude that for (near) axisymmetric galaxies the combination of
photometric and two-dimensional stellar kinematic observations can
strongly exclude triaxiality, but regular kinematics do not further
tighten the intrinsic flattening significantly. As a result, we expect
the inclination of fast-rotator early-type galaxies to remain nearly
unconstrained above the photometric lower limit, although better
observations do seem to slightly decrease the range in inclinations
(\S\ref{ssub:soblatemods}). Triaxial galaxies can have additional
complexity in both the observed photometry and kinematics, such as the
central KDCs observed in many slow-rotator early-type galaxies, which
allows the intrinsic shape to be accurately recovered. The intrinsic
shape of round galaxies with no significant rotation is degenerate,
unless additional constraints such as from a thin disk are available.

\section*{Acknowledgments}
\label{sec:acknowledgments}

We thank Anne-Marie Weijmans and Tim de Zeeuw for useful comments on the
manuscript, and Michele Cappelari for many lively discussions. RvdB
acknowledges support from grant 614.000.301 from NWO and the LKBF. GvdV
acknowledges support provided by NASA through Hubble Fellowship grant
HST-HF-01202.01-A awarded by the Space Telescope Science Institute, which is
operated by the Association of Universities for Research in Astronomy, Inc.,
for NASA, under contract NAS 5-26555. This paper uses software from
\cite{2002MNRAS.333..400C}, \cite{2003MNRAS.342..345C} and \cite*{2003galahad}.


\bibliographystyle{mn2e}


\bsp 

\label{lastpage}

\end{document}